\documentclass[fleqn,12pt,twoside]{article}
\usepackage[headings]{espcrc1}

\readRCS
$Id: espcrc1.tex,v 1.2 2004/02/24 11:22:11 spepping Exp $
\usepackage{graphicx}
\usepackage[figuresright]{rotating}

\newcommand{\AmS}{{\protect\the\textfont2
  A\kern-.1667em\lower.5ex\hbox{M}\kern-.125emS}}

\title{Probing Quark Gluon Liquid Using Transverse Momentum Fluctuations}

\author{Mohamed Abdel-Aziz\address[MCSD]{Department of Physics, Wayne State University,
        Detroit, MI 48201, USA}%
        \thanks{Current address: Institut f\"ur Theoretische Physik, J.W. Goethe Universit\"at,
        60438 Frankfurt am Main, Germany}
                 and
        Sean Gavin\addressmark[MCSD]\thanks{This work was supported in part by a U.S. National Science
foundation CAREER/PECASE award under grant PHY-0348559.}}

\runtitle{Probing Quark Gluon Liquid using Transverse Momentum
Fluctuations} \runauthor{S. Gavin}

\begin{document}

\maketitle

\begin{abstract}
The onset of equilibration of partons in nuclear collisions may lead
to related trends in the centrality dependence of  $\langle
p_t\rangle$, $p_t$ fluctuations, and net charge fluctuations
\cite{gavin}. We extend the transport description of
ref.~\cite{gavin} to include radial flow.
\end{abstract}

\section{INTRODUCTION}
Liquids exhibit shorter correlation lengths than gases. Correlation
and fluctuation measurements can therefore reveal the state --
liquid or gaseous -- of the matter produced in heavy ion collisions,
but we must first sort out the ion-collision effects that also
contribute. Experiments show that dynamic fluctuations of the
transverse momentum are roughly independent of beam energy, but
increase as centrality increases \cite{pruneau,CERES,STAR,PHENIX}.
Local equilibration can explain this behavior \cite{gavin}, but flow
and jet explanations have also been suggested \cite{pruneau,PHENIX}.
Certainly all effects play a role at some level.  It is therefore
important to describe all of them in a common framework. We take a
modest step in this direction by extending the formulation in
\cite{gavin} to account for contributions from radial flow.

Dynamic fluctuations are obtained from the measured
fluctuations by subtracting the statistical value expected in global
equilibrium \cite{pruneau}. For particles of momenta $\mathbf{p}_1$ and
$\mathbf{p}_2$, dynamic multiplicity fluctuations are characterized
by
\begin{equation}\label{eq:DynamicMult}
    R={{\langle N^2\rangle -\langle N\rangle^2 -\langle N\rangle}\over{\langle
    N\rangle^2}}={{1}\over{\langle N\rangle^{2}}}\int\! d\mathbf{p}_{1}d\mathbf{p}_{2}\,
    r(\mathbf{p}_{1},\mathbf{p}_{2}),
\end{equation}
where $\langle \cdots\rangle$ is the event average. This quantity
depends only on the two-body correlation function
$r(\mathbf{p}_{1},\mathbf{p}_{2}) = N(\mathbf{p}_{1},\mathbf{p}_{2})
- N(\mathbf{p}_1)N(\mathbf{p}_2)$. It is obtained from the
multiplicity variance by subtracting its Poisson value $\langle
N\rangle$. For dynamic $p_t$ fluctuations one similarly finds
\begin{equation}\label{eq:Dynamic}
    \langle \delta p_{t1}\delta p_{t2}\rangle =
    \int\! d\mathbf{p}_{1}d\mathbf{p}_{2}\,
    {{r(\mathbf{p}_{1},\mathbf{p}_{2})}\over{\langle
    N(N-1)\rangle}}
    \delta p_{t1} \delta p_{t2},
\end{equation}
where $\delta p_{ti} = p_{ti}-\langle p_t\rangle$. CERES and STAR measure
this observable, while PHENIX measures
$F_{p_t} \approx N \langle\delta p_{t1}\delta
p_{t2}\rangle/2\sigma^2$, where $\sigma^2 = \langle
p_t^2\rangle - \langle p_t\rangle^2$.

\begin{figure}[htb]
\begin{minipage}[t]{80mm}
\includegraphics[width=3.0in]{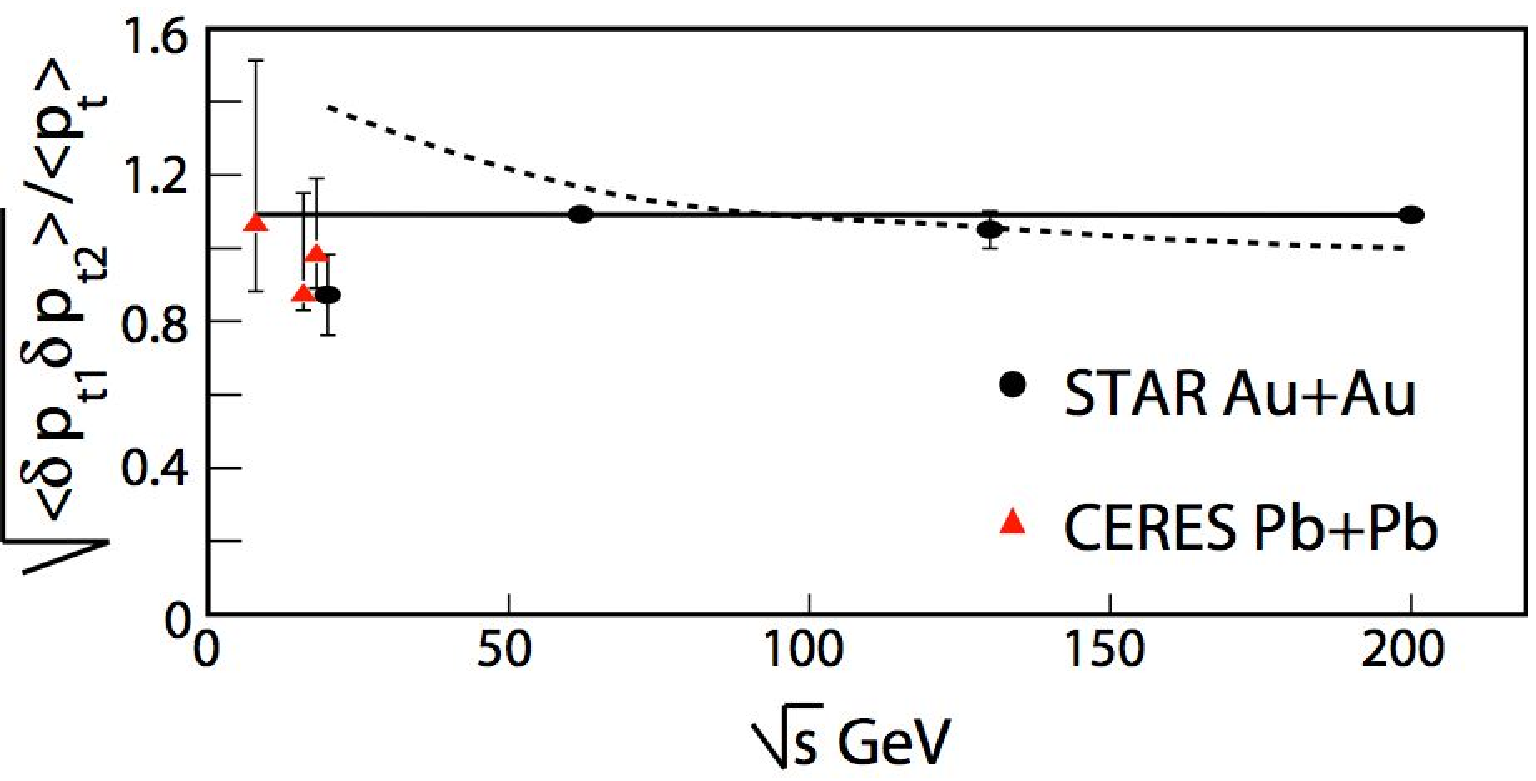}
\caption{Energy dependence of $p_t$ fluctuations from
\cite{pruneau}. CERES and STAR data are from \cite{CERES,STAR}. The
solid line follows from (\ref{eq:near}).} \label{fig:fig1}
\end{minipage}
\hspace{\fill}
\begin{minipage}[t]{75mm}
\includegraphics[width=3.2in]{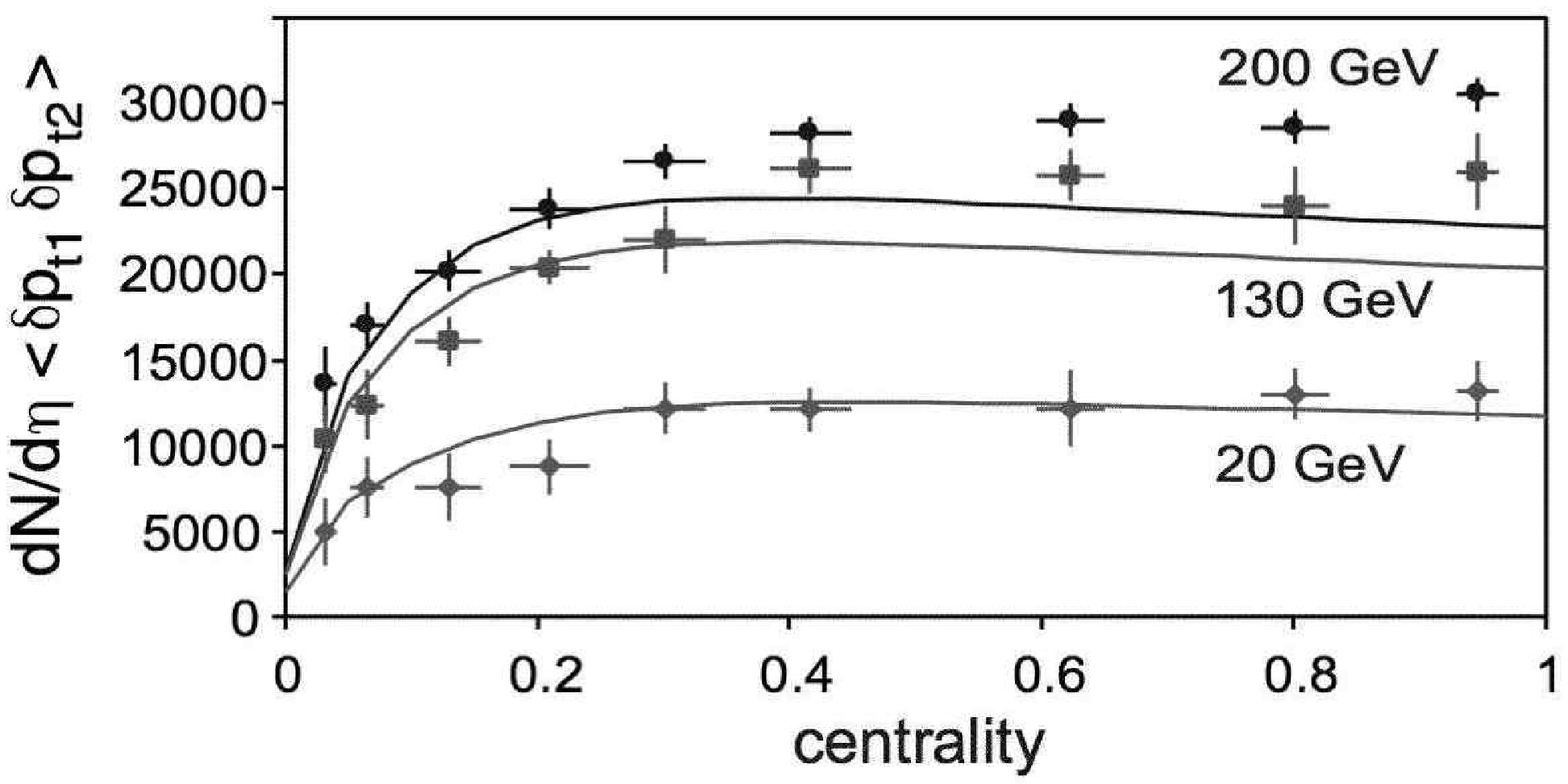}
\caption{Centrality dependence of $p_t$ fluctuations \cite{STAR}.
Centrality is determined by the number of participants relative to
the maximum value.} \label{fig:fig2}
\end{minipage}
\end{figure}

Figure 1 shows the ratio $\langle \delta p_{t1}\delta
p_{t2}\rangle^{1/2}/\langle p_t\rangle$ for central collisions as a
function of beam energy as compiled in \cite{pruneau} from CERES and
STAR data \cite{CERES,STAR}. The data suggest that the source of
fluctuations is independent of energy, although uncertainties are
large. This energy independence suggests that jets are not the leading
contribution to $p_t$ fluctuations.

\section{THERMALIZATION}

In \cite{gavin}, the Boltzmann transport equation is used to show
that thermalization alters the average transverse momentum according
to
\begin{equation}\label{eq:meanPt}
    \langle p_t\rangle = \langle p_t\rangle_o S + \langle
    p_t\rangle_e (1-S),
\end{equation}
where $S\equiv e^{-{\mathcal N}}$ is the probability that a particle
escapes the collision volume without scattering. The initial value
$\langle p_t\rangle_{o}$ is determined by the particle production
mechanism. If the number of collisions $\mathcal{N}$ is small,
$S\approx 1 -\mathcal{N}$ implies a random-walk-like increase of
$\langle p_t\rangle$ relative to $\langle p_t\rangle_{o}$. As
centrality increases, the system lifetime and $\mathcal N$ both
increase, eventually to a point where local equilibrium is reached.
Correspondingly, the survival probability $S$ in (\ref{eq:meanPt})
decreases with increasing centrality. The average $\langle
p_t\rangle$ increases for more central collisions to the point where
local equilibrium is established. The behavior in events beyond that
centrality depends on how the subsequent hydrodynamic evolution
changes $\langle p_t\rangle_e$ as the lifetime increases, as
discussed in the next section.

Dynamic fluctuations depend on two-body correlations and,
correspondingly, are quadratic in $S$. In \cite{gavin},
Langevin noise is added to the Boltzmann equation to describe the
fluctuations of the phase space distribution. If the initial
correlations are not far from the local equilibrium value, we find
\begin{equation}\label{eq:ptFluct}
    \langle \delta p_{t1}\delta p_{t2}\rangle
    =\langle \delta p_{t1}\delta p_{t2}\rangle_o S^2
    + \langle \delta p_{t1}\delta p_{t2}\rangle_e (1-S^2).
\end{equation}
As before, the initial quantity $\langle \delta p_{t1}\delta
p_{t2}\rangle_o$ is determined by the particle production mechanism,
while $\langle \delta p_{t1}\delta p_{t2}\rangle_e$ describes the
system near local equilibrium.

Near local equilibrium, spatial correlations occur because the fluid
is inhomogeneous -- it is more likely to find particles where the density
is high. The mean $p_t$ at each point is proportional to the temperature
$T(\mathbf{x})$, so that
\begin{equation}\label{eq:dTdT}
\langle \delta p_{t1}\delta p_{t2}\rangle_e\sim \int
r(\mathbf{x}_1,\mathbf{x}_2) {\delta T}(\mathbf{x}_1) {\delta
T}(\mathbf{x}_2),
\end{equation}
where $r(\mathbf{x}_1,\mathbf{x}_2)$ is the spatial correlation
function and $\delta T = T - \langle T\rangle$.  Our comparison to
the STAR data computed in \cite{gavin} is shown in fig.~2.

To see how the energy dependence of $\langle \delta p_{t1}\delta
p_{t2}\rangle^{1/2}/\langle p_t\rangle$ from (\ref{eq:dTdT})
compares to the data, observe that \cite{gavin} finds
\begin{equation}\label{eq:near}
    \langle \delta p_{t1}\delta p_{t2}\rangle_{e}
    = F{{\langle p_t\rangle^2R}\over{ 1+R}},
\end{equation}
where $R$ is the scaled variance (\ref{eq:DynamicMult}). The
dimensionless quantity $F$ depends on the ratio of the correlation
length to the transverse size. In \cite{gavin}, we took $F$ and $R$
to be energy independent. Therefore, if  matter is locally
equilibrated in the most central collisions, we would obtain the
solid line in fig.~1.

We do not expect $F$ to vary strongly with energy unless the
correlation length changes, e.g., from a liquid to a gas value. In
addition, we expect the scaled variance $R$ to vary primarily with
the number of participants and, hence, to be roughly independent of
energy.  For comparison, the dashed curve in fig.~1 shows an ad hoc
$N^{-1/2}$ dependence on the measured multiplicity.

\section{FLOW}
Thermalization describes the $p_t$ fluctuations measured by STAR
\cite{STAR} in
peripheral collisions as shown in fig.~2, but model calculations
fall short of the data for the most central collisions. One
explanation of this discrepancy is the appearance of radial collective flow
in the nearly-thermalized central collisions. Radial flow  was
omitted in \cite{gavin}, which only addressed longitudinal expansion.
Flow arises when pressure gradients within
the fluid cause it to move outward with an average local velocity.
This collective motion requires the fluid to be sufficiently near
local equilibrium that each fluid cell can respond to the different
pressures of nearby cells.

To see how flow fits into our transport framework, consider a test
particle traveling through the medium. As the system begins to
thermalize, the test particle undergoes a random walk so that its
root-mean-square $p_t$ increases. At the same time, all of the other
particles are experiencing their own random walks in both position
and momentum space. On average, the walk in position space directs
particles radially outward with an average drift velocity $v$,
taking along the test particle. Collective flow builds as more of
the particles encountered by the test particle are moving radially
outward. We expect this effect to be most important in central
collisions, since the system lives longer and particles have more
time to interact.
\begin{figure}[htb]
\begin{minipage}[t]{80mm}
\includegraphics[width=3.0in]{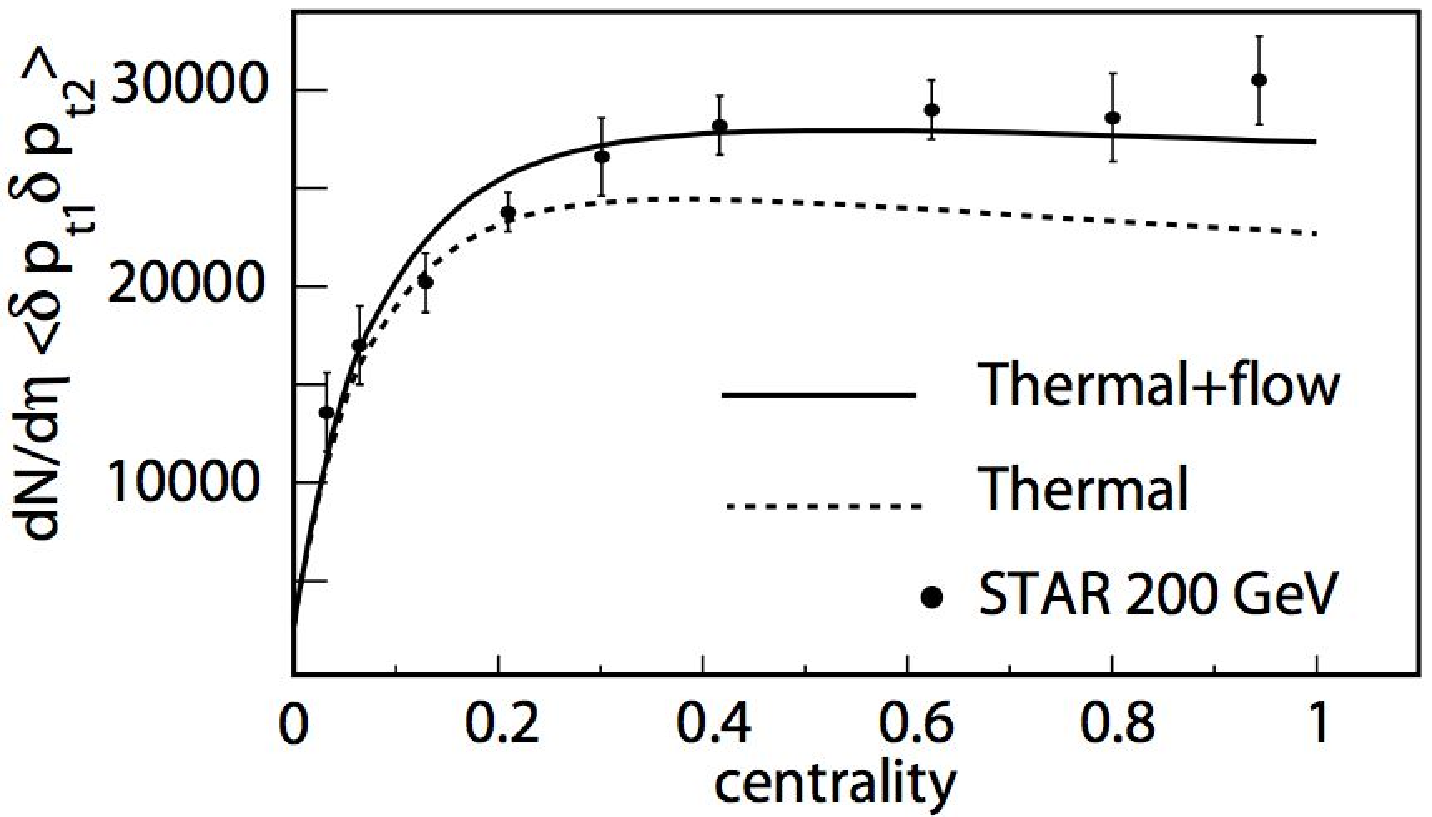}
\caption{Centrality dependence of $p_t$ fluctuations with radial
flow included. } \label{fig:fig3}
\end{minipage}
\hspace{\fill}
\begin{minipage}[t]{75mm}
\includegraphics[width=2.6in]{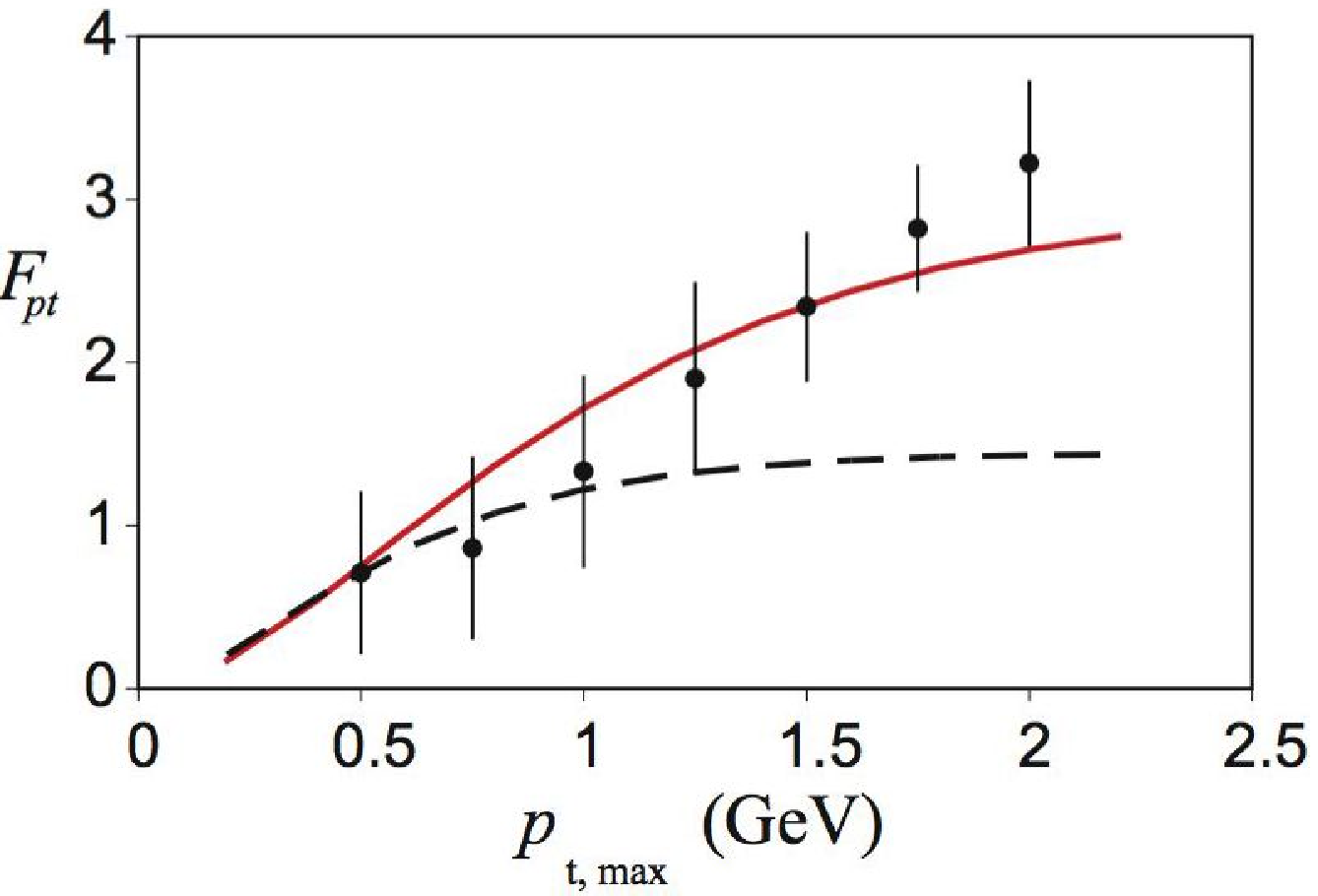}
\caption{PHENIX data for central Au+Au for $p_t < p_{t,\,{\rm max}}$
\cite{PHENIX}. The solid curve includes flow, while dashed curve
does not.} \label{fig:fig4}
\end{minipage}
\end{figure}

A simple way to describe flow is to introduce an average radial
parton drift velocity $v$.  This drift causes a radial blue-shift
for massless partons, so that the parton
Boltzmann distribution is $f(\mathbf{p})=\exp(-E/T_{\rm eff})$,
where $T_{\rm eff}=T[(1+v)/(1-v)]^{1/2}$. The average $p_t$ and its fluctuations
near equilibrium then satisfy
\begin{equation}\label{eq:ptDrift}
\langle p_t\rangle_e
\propto
\left(\frac{1+v}{1-v}\right)^{1/2}
\,\,\,\,\,\,\,\,\,\,\,\,\,\,\,\,\,\,\,\,{\rm and}
\,\,\,\,\,\,\,\,\,\,\,\,\,\,\,\,\,\,\,\, \langle \delta p_{t1}\delta
p_{t2}\rangle_{e}
\propto
 \left({\frac{1+v}{1-v}}\right).
\end{equation}
%
Observe that $\langle \delta p_{t1}\delta
p_{t2}\rangle^{1/2}/\langle p_t\rangle$ is independent of the flow
velocity $v$; fig.~1 is unaffected.
We obtain the mean velocity as a function of centrality by fitting
the $\langle p_t\rangle$ measured by PHENIX for pions while using
the same $\langle p_t\rangle^o_e$ as before \cite{thesis}. The net
effect of including velocity in (\ref{eq:ptDrift}) is to compensate
the longitudinal cooling.

By including flow we describe the data for the full centrality range,
see fig.~3.  Flow effects are further exhibited by measuring fluctuations
in the acceptance region $0.2$~GeV~$< p_t < p_{t,\,{\rm
max}}$ as a function of $p_{t,\,{\rm max}}$ following PHENIX \cite{PHENIX}.
Observe that (\ref{eq:near}) implies $F_{p_t} \approx N \langle\delta
p_{t1}\delta p_{t2}\rangle/2\sigma^2 \propto NR\langle
p_t\rangle^2/\sigma^2\sim NR$. The multiplicity $N$ grows strongly as
$p_{t,\,{\rm max}}$ is increased, since $N$ is the integral of
$\exp(-p_t/T_{\rm eff})$ over the acceptance. To see that $R$ is
independent of $p_{t,\,{\rm max}}$, let $\epsilon$ be the
probability that a particle falls in the acceptance region and
$1-\epsilon$ be the probability that it is missed. The average
number of detected particles is $\langle N\rangle_\epsilon =
\epsilon\langle N\rangle$. For a binomial distribution, one has $\langle
N^2\rangle_\epsilon = \epsilon^2\langle
N^2\rangle+\epsilon(1-\epsilon)\langle N\rangle$. Equation
(\ref{eq:DynamicMult}) then implies that $R(\epsilon) = R$, the
value for full acceptance.

The dashed curve in fig.~\ref{fig:fig4} is computed from
(\ref{eq:near}) at a freeze out temperature fixed by the measured
$\langle p_t\rangle$. The solid curve includes flow through $T_{\rm
eff}$ and fits the data much better.

In summary, we find that flow is an important aspect of the multiple
scattering contribution to $p_t$ fluctuations in central collisions.
The flow and thermalization contributions are essentially
independent of beam energy in accord with data \cite{pruneau}. Any
jet contribution to this phenomena would be tightly constrained by
these data \cite{thesis}.

\end{document}